\documentclass[12pt,preprint]{aastex}

\shorttitle{MID-INFRARED OBSERVATIONS OF GD 1400}
\shortauthors{J. Farihi}

\begin{document}

\title{MID-INFRARED OBSERVATIONS OF THE WHITE DWARF
	BROWN DWARF BINARY GD 1400}

\author{J. Farihi\altaffilmark{1,2},
	 B. Zuckerman\altaffilmark{2}, \& 
	 E. E. Becklin\altaffilmark{2}}

\altaffiltext{1}{Gemini Observatory,
			Northern Operations,
			670 North A'ohoku Place,
			Hilo, HI 96720; jfarihi@gemini.edu}
\altaffiltext{2}{Department of Physics \& Astronomy,
			University of California,
			430 Portola Plaza,
			Los Angeles, CA 90095}



\begin{abstract}

Fluxes are measured for the DA white dwarf plus brown dwarf pair
GD 1400 with the Infrared Array Camera on the {\em Spitzer Space
Telescope}.  GD 1400 displays an infrared excess over the entire
$3-8\mu$m region consistent with the presence of a mid- to late-type L
dwarf companion.  A discussion is given regarding current knowledge
of this unique system.

\end{abstract}


\section{INTRODUCTION}

White dwarfs make excellent infrared targets with
favorable contrast between cool self-luminous orbiting
bodies and target star.  This fact was perhaps first 
appreciated by \citet{pro82} and later by others who 
realized the potential for the detection of substellar
companions \citep{pro83,shi86,zuc87,zuc92}.  Although low
mass stars and brown dwarfs have cool effective temperatures
and are intrinsically faint, their radii ($R\approx1R_J$;
\citealt{bur97,cha00,cha00b,bur01}) are approximately 10
times larger than white dwarf radii ($R\approx1R_{\oplus}$;
\citealt{ber95a,ber95b}).  Therefore, spatially unresolved
low mass stellar and substellar companions to white dwarfs
can be detected in the infrared as excess emission.  These
facts led to the discovery of the first brown dwarf candidate,
the prototype L dwarf, and the coolest known ``star'' for about
7 years, GD 165B \citep{bec88}.  Currently, GD 165 and GD
1400 are the only two white dwarfs known to
have L-type
companions; these are possible or likely brown dwarfs,
respectively \citep{kir99,far04b}.

It is clear from large ($N>350$) surveys that L dwarf
companions to white dwarfs are not common, with frequency
$\la0.5$\% \citep{far04,far05}.  While it is true that
cooling brown dwarfs pass through the more luminous M \&
L dwarf stages faster than later stages, the lowest mass
stars (0.075 $M_{\odot}\la M <0.10$ $M_{\odot}$) should have
spectral types in the range M6$-$L3 at white dwarf ages
and beyond \citep{cha00,cha00b,bur01}.  Hence the relative
dearth of both L and late M dwarf companions to white dwarfs,
compared to the frequency of earlier M-type companions and to
the number of known L and late M dwarfs in the field \citep{zuc92,
gre00,wac03,far04,far05}, is indicative of binary star formation
-- specifically low mass companion formation to intermediate mass
stars -- and not a limitation or bias in searches.

Relative to the sensitivity and wide
range of
separations that have been probed for L dwarfs,
there is
less evidence against the presence of T dwarf and later
type
companions to white dwarfs \citep{far04,far05,dob05}.
Both H$_2$O \& CH$_4$ absorption, plus H$_2$ collision-induced
absorptions suppress some regions of near-infrared flux in T
dwarfs \citep{bur02}, causing them to appear blue in the $J-K$
color index.  This makes T dwarf companions photometrically
undetectable as excess emission at $2.2\mu$m, unless the white
dwarf primary has $M_K\ga13$ mag \citep{far04,far05}, corresponding
to very cool and/or very massive degenerates ($T_{\rm{eff}}<7000$ K
for log $g=8.0$, or $T_{\rm{eff}}<9000$ K for log $g=8.5$; \citealt
{ber95a,ber95b}).  Although the majority of known white dwarfs do
not meet these faintness criteria \citep{mcc99}, many nearby cool
degenerates have been photometrically surveyed in the near-infrared
for various purposes, with no evidence of T dwarf secondaries via
$K$ band excess \citep{ber97,leg98,ber01,far04,far05}.  Owing to 
these facts, generally speaking, T-type and later companions to
white dwarfs are only detectable in the near-infrared as spatially
resolved objects in deep ground-based or space-based imaging, or
as unresolved secondaries with high S/N spectroscopy \citep{burl02,
far04,far04b,far05,dob05}.  For wavelengths $\ga3\mu$m, the contrast
is once again favorable for the detection of the coolest brown dwarf
companions \citep{bur03}.  However, due to large thermal background,
there are very few white dwarfs bright enough to be observed from
the ground at $3\mu$m and beyond, where companions later than L-type
might be detectable around a significant fraction of known degenerates.
Hence space-based imaging and spectroscopy at $\lambda\ga3\mu$m with
{\em Spitzer} is currently the only way to survey a large number of
white dwarfs for unresolved T-type and later brown dwarf companions.

This paper presents mid-infrared fluxes and magnitudes for
GD 1400 (DA4.3+dL6.5, $d\approx39$ pc; \citealt{far04b,dob05})
measured with the Infrared Array Camera (IRAC; \citealt{faz04})
on {\em Spitzer}.  The IRAC data are found to be consistent
with expectations based on previously published ground-based
near-infrared photometric and spectroscopic data.

\section{DATA \& ANALYSIS}

GD 1400 was observed with IRAC in all four channels as
part of a program
searching for substellar companions
to
nearby white dwarfs.  The imaging strategy consisted of
30 second frame times in a 20-point, medium-scale (median
move of $\sim40''$) cycling dither pattern.  In this way
the point spread function is well sampled, saturation of
target and candidate companions is avoided, effects of
cosmic rays, detector blemishes, and bad pixels can be
removed, flat fielding errors are minimized, and detector
artifacts due to bright sources on and off the chip are
more easily eliminated.  Thus, a total integration time
of 600 seconds was achieved at each of these wavelengths;
3.6, 4.5, 5.7, \& 7.9$\mu$m \citep{ssc05}.

The individual frames were combined into a single
image
via the IRAC calibration pipeline, version 11.0.  For single
images, the basic calibrated data contain dark and sky subtraction,
linearization, flat-fielding, cosmic ray detection, flagging of bad
pixels, and flux calibration.  The final processed image includes
pointing refinement, mosaicking (image registration onto a larger
grid), masking, and coaddition with outlier rejection \citep{ssc05}.
Aperture photometry was performed on the target in the final combined
image using standard
IRAF tasks.  Both the flux and signal-to-noise
(S/N) were measured in a
$2-3$ (wavelength dependent) pixel radius,
with a sky annulus of $r=10-20$ pixels.  This measured flux was then
corrected to the standard 10 pixel aperture radius using aperture
corrections found in \citet{ssc05}.  The results are listed in
Table \ref{tbl-1}.

\section{RESULTS \& DISCUSSION}

\subsection{$2-8\mu$m Colors}

Little is known about the probable white dwarf
plus
brown dwarf spectroscopic binary, GD 1400.
The cool companion was discovered,
then confirmed,
through photometric excess and subsequent
spectroscopy
in the $2.2\mu$m region \citep{far04b}.  To give a brief
summary; its apparent lack of excess emission at $1.2\mu$m
implies that GD 1400B has a spectral type of L5.5 or later
and the lack of Na in its $K$ band spectrum indicates it
cannot be an early L dwarf.  Utilizing the best available
data on the white dwarf primary to assess its distance and
to account for its contribution at near-infrared wavelengths,
the absolute magnitude of GD 1400B would place it around
spectral type L6 \citep{far04b}.  Subsequently, an independent
spectroscopic study estimated GD 1400B at spectral
type L7 through simultaneous fits of the white dwarf
and brown dwarf components in an HK grism observation,
with model and empirical template spectra respectively
\citep{dob05}.

\begin{figure}

\plotone{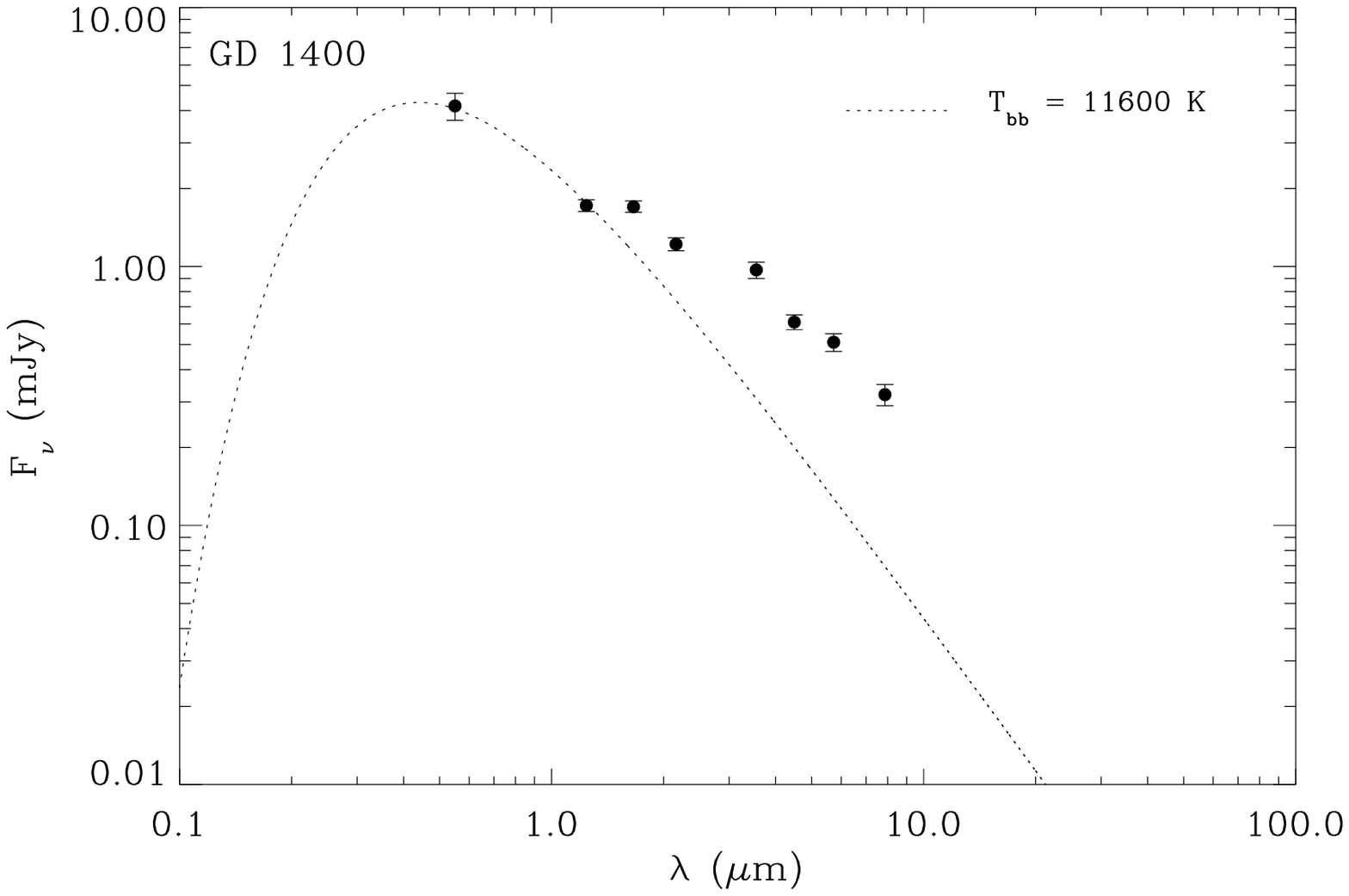}
\caption{Spectral energy distribution of GD 1400, demonstrating
the presence of the spatially unresolved cool brown dwarf companion.
Optical and near-infrared ($VJHK$) data are from \citet{far04b}.
\label{fig1}}
\end{figure}

The IRAC measurements of GD 1400 presented
in Figure \ref{fig1} \& Table \ref{tbl-1} have
${\rm S/N}>15$ at all wavelengths.  In Table 
\ref{tbl-2}, the expected flux from GD 1400A at
IRAC wavelengths has been calculated from the data
in \citet{far04b}, then subtracted from the total flux
to produce the contribution of GD 1400B, with errors.
It is noteworthy that using the Rayleigh-Jeans approximation
longward of $2\mu$m does not yield zero color as expected
for an 11,600 K star.  This is a consequence of the IRAC
zero magnitude flux scale \citep{ssc05} and not a reflection
of any intrinsic property of GD 1400A.  For completeness,
Table \ref{tbl-2} lists the resulting deconvolved magnitudes
for GD 1400B using both Rayleigh-Jeans and zero color assumptions
for GD 1400A.  Table \ref{tbl-3} lists the $2-8\mu$m colors implied
by the Table \ref{tbl-2} magnitudes for GD 1400B, which are minimally
affected by the choice of model for GD1400A.  The deconvolved
magnitudes of the cool companion imply near- to mid-infrared
colors consistent with those measured for isolated mid to late
L dwarfs \citep{pat04}.  The last column of Table \ref{tbl-3}
lists the ranges of L dwarf types consistent with each color
index, based on relations in \citet{pat04}.  The $2-8\mu$m
colors of GD 1400B are consistent with a spectral type of
L5$-$L7, corroborating previous findings by alternate
methods.

Looking at Figure \ref{fig1}, there appears to
be a relative drop in flux at $4.5\mu$m compared
to the other three IRAC bandpass measurements.
Accordingly, the [3.6]$-$[4.5] color index of GD
1400B is bluer than its other IRAC colors, and is
actually just negative.  A slight negative [3.6]$
-$[4.5] color index is also seen in the majority of
all L dwarfs in \citet{pat04}, but appears to be a
bit more pronounced in types L5 and later.  The 
reason for this is probably the presence of a wide
CO absorption feature that spans a decent portion
of the $4.5\mu$m IRAC bandpass, and which becomes
more pronounced at later L dwarf types \citep{sau03}.
Although no currently published L dwarf spectra span
the entire $3-5\mu$m range, some ground-based $L$ band
spectra of L dwarfs exist, and the beginnings of this 
CO feature may be what is seen to cause a drop in flux
just before $4\mu$m \citep{cus05}.

The remaining IRAC data points in Figure
\ref{fig1} do not deviate drastically from
a Rayleigh-Jeans type slope, as seen in M dwarfs
\citep{roe04,cus05}.  Hence, it is unlikely that
GD 1400B is cool enough to have formed a significant
amount of CH$_4$ yet, as L8$-$L9 spectral types have
\citep{geb02,cus05}, which can cause IRAC colors to
more nearly resemble those of T dwarfs \citep{pat04}.

\subsection{Origin \& Evolution}

Because GD 1400AB has yet to be spatially
resolved ($a\la0.3''$; \citealt{far04b}, it
remains possible that this spectroscopic binary
is a radial velocity variable.  It is perhaps more
likely the system resides in close orbit due to the
fact that post-asymptotic giant branch (AGB) evolution
predicts a bimodal distribution of orbital semimajor
axes for low mass, unevolved companions to white dwarfs
\citep{far04}.  Specifically, companions close enough
to orbit within the AGB envelope should spiral inward
due to transfer of orbital energy into the envelope
via friction \citep{pac76,liv84,liv88}, while those
outside the envelope should spiral outward due to
weakened gravity from mass loss \citep{jea24,zuc87,
burl02,far04}.  It is not known exactly where the critical
radius lies for inward versus outward orbital alteration,
and it must depend on companion mass, but a reasonable
assumption is on the order of a few AU based both on
theory which includes tidal interactions (important
for the lowest mass companions) and the fact that
intermediate mass stars should evolve to have AGB
photospheres $\sim1-2$ AU in radius \citep{sac93,
ras96,dun98,sie99,pas01,burl02}.  Observations
separated by 2 days in 2000 July revealed a small,
$\sim4$\% variation in radial velocity (R. Napitwotzki
2005, private communication).  The measured
variation is not inconsistent with a very low mass
($\sim0.06$ $M_{\odot}$) companion with an orbital
period greater than several days. 

Further radial velocity monitoring of the white
dwarf in the optical and/or its companion in the
near-infrared, or high resolution ground- or
space-based imaging should eventually reveal
the nature of the current orbital separation
of the binary.  Resolving the pair would be
advantageous because the companion could be directly
studied.  On the other hand, it would be fortuitous
if the system were a radial velocity variable because
then the mass and radius of the secondary could be
estimated.  Currently, there is only a single L dwarf
(binary) system with a mass measurement \citep{bou04},
and no mass estimates for old brown dwarfs.  There
exist two independent and reliable spectroscopic
fits of $T_{\rm eff}$ and log $g$ for GD 1400A,
and hence the mass of the white dwarf is fairly
well constrained near $M\approx0.7$ $M_{\odot}$
\citep{koe01,fon03}.  A trigonometric parallax
and high precision optical photometry would
tighten up the primary mass estimate, making
any secondary mass determination more reliable.

Determining the orbital parameters of this
so far unique binary is critical to understanding
the origin and evolution of the brown dwarf secondary.
It is likely that the system formed as an extreme low
mass ratio binary ($M_2/M_1\approx0.02$; \citealt{far04b}),
but it is conceivable that the companion formed in a massive
disk around a $\sim3$ $M_{\odot}$ main sequence star.
There have been several substellar companions detected 
around K giants \citep{fri02,mit03}, which are the
descendents of main sequence A \& F stars.  Presumably,
these substellar companions formed in their respective
primary progenitor disks based on their current orbital
semimajor axes.  Will these brown dwarfs survive the
current first-ascent and ensuing asymptotic giant
branches to become companion systems similar to
GD 1400?  Although complete evaporation or inspiral
collision with the stellar core is possible inside
the AGB envelope, the higher mass brown dwarfs around
these K giants may persist, as has GD 1400B, either by
eschewing the greatly expanded photosphere or simply
surviving the envelope itself \citep{liv84,ibe93,sac93,
ras96,dun98,sie99,burl02}.

It is not known whether GD 1400AB shared a common
envelope during the AGB phase of the primary, but
hopefully more data will soon give indications one 
way or another.  Important questions regarding this
stage of evolution and its outcome are: (1) was any
mass accreted by the low mass companion?; (2) what
were the initial masses and separation of the binary?;
(3) is the pair close enough now to interact in any
way which might be detectable?  Ultimately, the core
science is to understand the origin of the current
binary stellar parameters, especially those of GD
1400B, and to discover in what ways these are
products of formation, evolution, or both.

At present, there is no hard evidence that low
mass unevolved companions ($M\la0.3$ $M_{\odot}$)
to white dwarfs in detached, post-common envelope
binaries have accreted a significant amount of mass
($\sim0.01-0.1$ $M_{\odot}$; \citealt{max98,cha00,far04,
far05}.  However, there is some tentative and indirect
evidence of secondary accretion from either the common
envelope or the more distant AGB wind in dwarf and giant
K stars, such as overabundances of carbon and $s$-process
elements and oversized apparent radii \citep{pol94,jef96,
bon03,dra03}.  First, it is uncertain whether these
observations imply accretion of an order which would
be noticeable in the mass distribution of post-common
envelope versus widely separated low mass companions to
white dwarfs; i.e. enough mass to transform a brown dwarf
into a star \citep{far04}.  Second, there are a few alternative
explanations for the aforementioned observations.  Accretion
from the AGB envelope or wind should take place more readily
for relatively more massive (K dwarf) or compact secondaries,
whereas companions with oversized apparent radii can be
explained by either irradiation from very hot helium burning
stars or starspots \citep{pol94,obr01}.  It is also possible
that a currently detached, post-common envelope binary was
previously in a semi-detached configuration, where mass
exchange might occur.  This can take place if enough mass
is lost adiabatically from the system to expand the orbit
\citep{jea24,nel01}.  There is still signifiant interest
in this issue and tests have been proposed to look for
evidence of secondary accretion from the envelope
\citep{sar95,dhi02}.

Another issue important for DA white dwarfs like GD 1400A
is possible atmospheric pollution by winds from close, low
mass companions.  In searches for metal-rich DA (DAZ) stars,
white dwarfs known to have unresolved red dwarf companions
are found to have Ca in their photospheres with significantly
greater frequency (60\%) than other types of double or single
white dwarf systems ($6-20$\%) \citep{zuc98,zuc03}.  Unseen
companions or planetary material have often been considered
a possible explanation both in particular instances and in
general for the DAZ phenomenon \citep{hol97,jur03,dob05}.
GD 1400A, observed for the SPY project \citep{koe01},
was spectroscopically searched for Ca H \& K lines, which were
not detected \citep{koe05}.  Based on the published sensitivity,
this should firmly rule out Ca abundances greater than [Ca/H]
$\approx-9$ (see Figure 2 of \citealt{koe05}).  If the pair is
close now, then there appears to be little if any Ca pollution
of GD 1400A by any wind from GD 1400B.

\section{CONCLUSION}

The mid-infrared data from {\em Spitzer}/IRAC corroborate
the presence of an apparently ordinary L dwarf companion to
GD 1400.  Additionally, the colors appear to agree with previous
determinations that place the companion safely into the substellar
regime, with a spectral type later than L5.  So far there is no
hard observational evidence to confirm or rule out a close orbit,
but this scenario appears more likely based on theoretical 
predictions and the available data.

Ongoing and future {\em Spitzer} observations can shed some
light on the frequency of ultracool low mass companions to white
dwarf stars, both as unresolved secondaries and widely separated
companions (Farihi, Zuckerman, \& Becklin, in preparation).

\acknowledgments

This work is based on observations made with the Spitzer
Space Telescope, which is operated by the Jet Propulsion
Laboratory, California Institute of Technology under NASA
contract 1407.  Support for this work was provided by NASA
through Contract Number 1264491 issued by JPL/Caltech.

\clearpage

\clearpage

\begin{deluxetable}{cccccc}
\tablecaption{IRAC Fluxes for GD 1400 \label{tbl-1}}
\tablewidth{0pt}
\tablehead{
\colhead{Channel}					&
\colhead{$\lambda_{0}$\tablenotemark{\dag} ($\mu$m)}	&
\colhead{$\Delta\lambda$\tablenotemark{\dag} ($\mu$m)}	&
\colhead{$F_{\nu}$ ($\mu$Jy)}				&
\colhead{S/N}						&
\colhead{Magnitude}}

\startdata

1	&3.55	&0.68	&971	&440	&$13.65\pm0.06$\\
2	&4.49	&0.87	&608	&190	&$13.68\pm0.06$\\
3	&5.73	&1.25	&513	&31	&$13.37\pm0.07$\\
4	&7.87	&2.53	&321	&17	&$13.25\pm0.08$\\

\enddata

\tablecomments{Not included in the S/N is the absolute
calibration uncertainty in the IRAC instrument at the time of
the pipeline processing, which was around 6\% (M. Cohen 2005,
private communication; \citealt{coh04}.  This uncertainty is 
included in the magnitude column.}

\tablenotetext{\dag}{Values taken from the IRAC Data Handbook 2.0
\citep{ssc05}.}

\end{deluxetable}

\clearpage

\begin{deluxetable}{cccccc}
\tablecaption{Magnitudes for GD 1400A \& B \label{tbl-2}}
\tablewidth{0pt}
\tablehead{
\colhead{Object}	&
\colhead{$K_s$ (mag)}	&
\colhead{[3.6] (mag)}	&
\colhead{[4.5] (mag)}	&
\colhead{[5.7] (mag)}	&
\colhead{[7.9] (mag)}}	

\startdata

GD 1400A\tablenotemark{\dag}
		&$15.09\pm0.12$	&$15.22\pm0.12$	&$15.26\pm0.12$	&$15.30\pm0.12$	&$15.37\pm0.12$\\
GD 1400B	&$15.10\pm0.20$	&$13.94\pm0.10$	&$13.97\pm0.10$	&$13.57\pm0.10$	&$13.42\pm0.11$\\
		&		&		&		&		&\\

GD 1400A\tablenotemark{\ddag}
		&$15.09\pm0.12$	&$15.09\pm0.12$	&$15.09\pm0.12$	&$15.09\pm0.12$	&$15.09\pm0.12$\\
GD 1400B	&$15.10\pm0.20$	&$13.98\pm0.10$	&$14.03\pm0.10$	&$13.62\pm0.10$	&$13.47\pm0.11$\\

\enddata

\tablecomments{$K_s$ data taken from \citet{far04b}.}

\tablenotetext{\dag}{Flux of GD 1400A extrapolated to IRAC wavelengths
using the Rayleigh-Jeans approximation.}

\tablenotetext{\ddag}{Flux of GD 1400A extrapolated to IRAC wavelengths
assuming zero color.}

\end{deluxetable}

\clearpage

\begin{deluxetable}{cccc}
\tablecaption{$2-8\mu$m Colors for GD 1400B \label{tbl-3}}
\tablewidth{0pt}
\tablehead{
\colhead{Index}				&	
\colhead{Color 1}			&
\colhead{Color 2}			&
\colhead{Sp Type\tablenotemark{\dag}}}	
\startdata

$K_s-[3.6]$	&$+1.16$	&$+1.12$	&L5$-$L7\\
$[3.6]-[4.5]$	&$-0.03$	&$-0.05$	&L0$-$L8\\
$[4.5]-[5.7]$	&$+0.40$	&$+0.41$	&L5$-$L8\\
$[5.7]-[7.9]$	&$+0.15$	&$+0.15$	&L0$-$L7\\

\enddata

\tablecomments{Colors 1 \& 2 are those implied by the magnitudes in
rows 2 \& 4 of Table \ref{tbl-2} respectively.}

\tablenotetext{\dag}{Range of L dwarf types consistent with
each measured color index, based on relations in \citet{pat04}.}

\end{deluxetable}

\end{document}